\documentclass[12pt]{emulateapj}
\def\zbootes{$z$Bo\"otes\,}

\slugcomment{To Appear in ApJS}
\shorttitle{zBo\"otes : z-band Observations of the NDWFS Bo\"otes Field}
\shortauthors{Cool}

\begin{document}
\title{zBo\"otes : z-band Photometry in the NOAO Deep Wide-Field Survey
Bo\"otes Field}
\author{Richard J. Cool\altaffilmark{1} }

\altaffiltext{1}{Steward Observatory, 933 N Cherry Avenue, Tucson
AZ 85721;
rcool@as.arizona.edu}

\begin{abstract}
We present \zbootes, a new $z$-band photometric imaging campaign of
7.62 square degrees in the NOAO Deep Wide-Field Survey (NDWFS)
Bo\"otes field.
In this paper, all of the images for this survey are released as
well as the associated
catalogs.  The final \zbootes catalogs are complete (at the 50\%
level)
to 22.7 mag over 50\% of the field. With these depths,
the \zbootes images should be sensitive to
$L^*$ galaxies to $z\sim1$ over much of the survey area.
These data have several possible applications
including searching for and characterizing high-redshift quasars and
brown dwarfs and providing added constraints to photometric redshift
determinations of galaxies and active galaxies to moderate redshift.
The \zbootes imaging adds photometric data at a new wavelength to
the existing wealth of multi-wavelength observations of the NDWFS
Bo\"otes field.

\end{abstract}

\section{Introduction}

In recent years,  a number of multi-wavelength surveys have been
completed
in order to understand the evolution of the multi-wavelength
properties
of galaxies and active galactic nuclei (AGNs)
throughout cosmic history.   Deep observations spanning
from the ultraviolet to the radio are time consuming and obtaining
spectroscopic follow-up observations of cataloged galaxies and AGNs
requires many nights on the largest telescopes available.  Thus,
the area
covered by many of the deepest multi-wavelength surveys is fairly
small.
With the advent of new wide-field optical and near-infrared
imagers and
multi-object spectrographs, as well as superb new space facilities
such
as {\it GALEX} and {\it Spitzer}, the amount of the sky observed
 at all possible wavelengths is growing steadily.

One of the early deep, wide area, optical surveys, the NOAO Deep
Wide-Field
Survey \footnote[1]{http://www.archive.noao.edu/ndwfs \\ http://www.noao.edu/noao/noaodeep}
(NDWFS; Jannuzi et al. {\it in prep.}, Dey et al., {\it in prep.})
consists of two $\sim9$ square degree fields (the Bo\"otes and Cetus
fields)  with excellent optical ($B_W$, $R$, and $I$)
and near-infrared ($K_s$) photometry.  The NDWFS Bo\"otes field
has become
a popular target for many investigators and now has been observed
across the full electromagnetic spectrum.  Deep GALEX pointings
provide near and
far-ultraviolet photometry for the field; the NDWFS images consist of
optical and near-infrared coverage, and the  FLAMINGOS Extragalactic
Survey
\citep[FLAMEX;][]{elston2005}
 observed the field to deep limits in $J$ and $K_{\mbox{s}}$.
 {\it Spitzer} has imaged
 the Bo\"otes field with both IRAC \citep[Spitzer Shallow
 Survey;]{eisenhardt}
 and MIPS.  Radio observations include deep Westerbork
 observations at 1.4 GHz \citep[]{devries2002} and imaging by
 the Faint Images of the Radio Sky at Twenty-Centimeters survey
 \citep[FIRST;][]{becker1995}.   {\it Chandra} has observed the
 Bo\"otes field for 5
 ksec \citep[XBo\"otes;][]{murray2005,kenter2005}.  Optical
 spectroscopy for
 several highly complete samples of galaxies and AGN has been
 completed
 with the Hectospec multi-object spectrograph on the MMT as part
 of the
 AGN and Galaxy Evolution Survey (AGES; Kochanek in prep).  A small
 region in the Bo\"otes field was observed as part of the LALA survey
 to search for galaxies at very high redshifts \citep{rhoads2000}.
 These data sets have allowed searches for high-redshift quasars and
 low mass stars using mid-infrared selection techniques
 \citep{stern2006},
 studies of the quasar luminosity
 function \citep{brown2006,cool2006}, the clustering of high-redshift
 galaxies \citep{brown2005,stanford2005, rhoads2004,brown2003},
 Lyman alpha emitting galaxies \citep{wang2004, dey2005, dawson2004,
 rhoads2003}, the broad band properties of
 AGNs \citep{brand2006ii,brand2005,stern2006,stern2005}, the X-ray
 properties of AGNs \citep{brand2006,kollmeier2005},
 the spectral properties of infrared sources
 \citep{weedman2006,desai2006,khan2005,higdon2005} and many other
 topics.

In this paper, we present new $z$-band observations of 7.6 square
degrees
in the NDWFS Bo\"otes region.  These catalogs reach several magnitudes
deeper than the public imaging released by the Sloan Digital
Sky Survey
(SDSS) and will provide a useful intermediate photometric measurement
between the I band data from NDWFS and near-infrared photometry from
FLAMEX. We  release the catalogs and reduced images for public use.
Throughout this paper, all magnitudes are AB magnitudes
\citep{oke1974}.

\begin{figure}[!t]
\centering{\includegraphics[angle=0, width=3.5in]{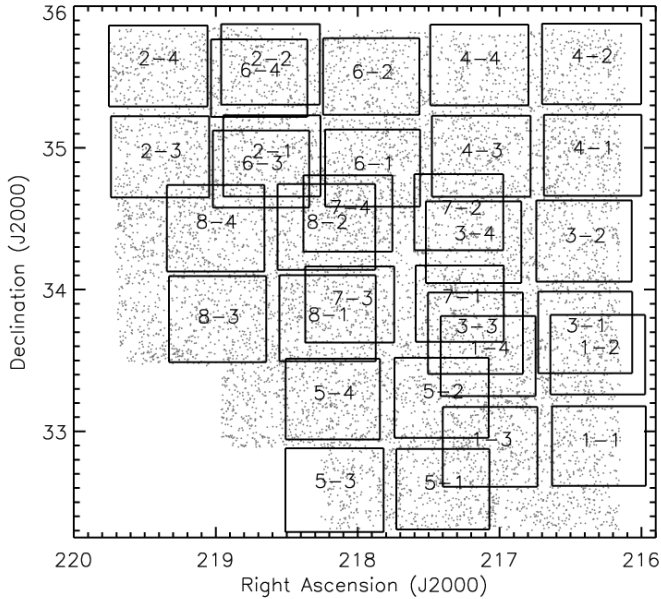}}
\caption{ \scriptsize Map of the NDWFS survey area  with the
coverage of the \zbootes catalog overlayed.  The grey points
illustrate
the distribution of 25\% of the $I<21.5$ galaxies in the NDWFS
catalogs. Each of the subfields of
the \zbootes imaging is labeled with the field number.  The \zbootes
imaging covers 7.6 square
degrees centered on the NDWFS Bo\"otes Field, providing another set of
data to the current suite of multi-wavelength observations
completed in
this region.}
\label{fig:fullcoverage}
\end{figure}

\section{Observations}

The \zbootes imaging survey was completed with the 90Prime
wide-field imager \citep{williams2004}
at prime focus  on the Bok 2.3m Telescope located on Kitt Peak.
This instrument
consists of four 4096x4096 thinned CCDs, providing excellent quantum
efficiency
in the blue, read out using eight amplifiers.   The chips are
arranged
in a windowpane
pattern with 10' gaps between each of the CCDs.  Each CCD images
a 30'x30'
field on the
sky with 0.45" pixels.  The south-east CCD has a large electron trap,
making 13\% of the area
on that chip unusable for photometric measurements.  These pixels are
masked throughout
the reduction process and are excluded when images are stacked.

The data were obtained between 28 Feb 2005 and 31 Mar 2005.
Sky conditions
varied from photometric to
moderate levels of cirrus throughout the observations. The typical
seeing during these observations was 1.6 arcsec. In total,
we completed
observations
for eight 90Prime pointings within the NDWFS Bo\"otes region.  At each
location, we obtained
several dithered 300s exposures (typically 12 exposures per field).
The telescope was moved
$\sim1'$ for each dither.    The total number of exposures for
each field
was determined by the conditions at the time of the observations.
All of the exposures for a single field were
obtained on the same night.   As the
gaps between CCDs on 90Prime are rather large, we did not attempt
to make
our dither pattern large enough the fill the regions between
each CCD.
This strategy avoids having non-uniform depth across a single
field but
 results in gaps in our photometric coverage
of the NDWFS survey field.  Figure \ref{fig:fullcoverage} shows the
region covered in the zBo\"otes survey compared to the NDWFS Bo\"otes
Survey
area. Throughout the rest of this paper, each of the fields imaged by
a single 90Prime CCD is treated independently.

\begin{figure*}[!t]
\centering{\includegraphics[angle=270, width=6in]{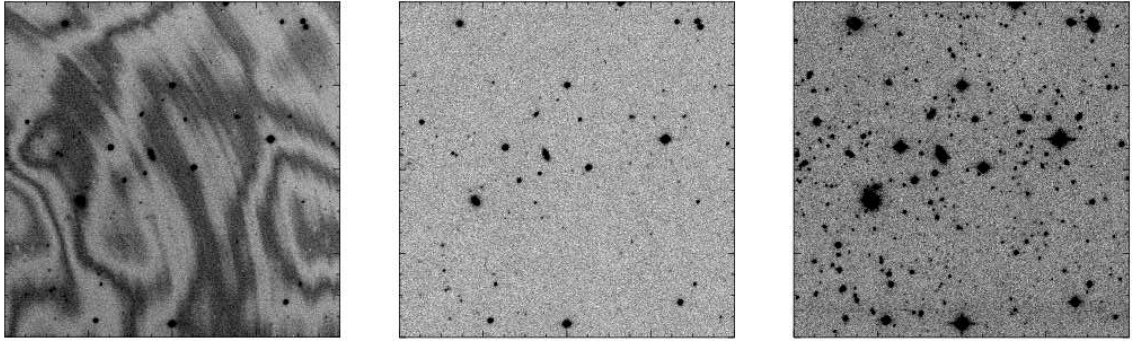}}
\caption{ \scriptsize Example of the reduction steps performed on the
\zbootes imaging data.  Frame (a) shows a portion of a raw data
frame  taken
with 90Prime.  Frame (b) illustrates the same frame after
flat-fielding
and fringe correction.  The final coadded frame, including  all
of the
observations for this field, is shown in frame (c).  Each of
the images
is 800 pixels on a side corresponding to $\sim6$ arcminutes. The
total exposure time for this field was one hour.}
\label{fig:reduction}
\end{figure*}

\section{Reductions}

All of the raw \zbootes images were processed using a combination
of home-grown IDL
routines and various
tasks available within the Image Reduction and Analysis Facility
(IRAF).
Each image was bias and overscan corrected and known bad columns
are removed by interpolating neighboring columns.   On each night
of clear
skies, observations of the twilight
sky were taken to generate a flat field image that was divided
into each
object frame to correct for pixel-to-pixel variations in the CCD
sensitivity.

While the thinned-chip nature of the CCDs on 90Prime allows for high
efficiency in the blue, it also results in strong
fringing in the reddest bands where the night sky spectrum is
dominated
by a forest of emission lines.  On each night of observation,
between 12
and 42 individual dithered images were obtained for this project
in the
$z$-band.  To generate a master fringe frame, we first removed
any large
scale gradient in the background of each input image.  The strength
of
the fringe pattern was then measured on each image and a
multiplicative
scale factor was applied to correct for any differences from
the mean.
We generated a master fringe frame by taking the median of all of the
individual images taken on a single night of observations.
This fringe frame was scaled to
match the average strength of the fringes in each individual exposure
and subtracted.  This process was iterated (typically twice) until
the fringe pattern was no longer present in each individual exposure.
Figure \ref{fig:reduction} (a) and (b) shows a region of a
single exposure before
any processing and after the fringe pattern was been removed.

The astrometry of each image was calibrated by locating stars
with $17<z<19$ from the Sloan Digital Sky Survey (SDSS) DR4
\citep{am2006}.
We fit the astrometric
solution with a 5th order TNX world-coordinate system (WCS) using the
IRAF task CCMAP.    The images were then de-projected onto a
rectilinear
pixel system using the task MSCIMAGE in the MSCRED package in IRAF.
Finally, aperture photometry of several stars in the magnitude
range $17<z<19$
was performed on each of the input frames in order to determine
offsets in the background level and photometric zeropoint between
each of
the frames.  Any variations in the mean background or transparency
were corrected
before the individual images were stacked to create the final
coadded frame for
each field.  Pixels with values more than $3\sigma$ from
the mean were
clipped when creating the final mosaic;  this clipping rejected
any cosmic
rays present in the individual frames.
Figure \ref{fig:reduction} shows a portion of a final stacked image.

\begin{figure}[b]
\centering{\includegraphics[angle=0, width=3.5in]{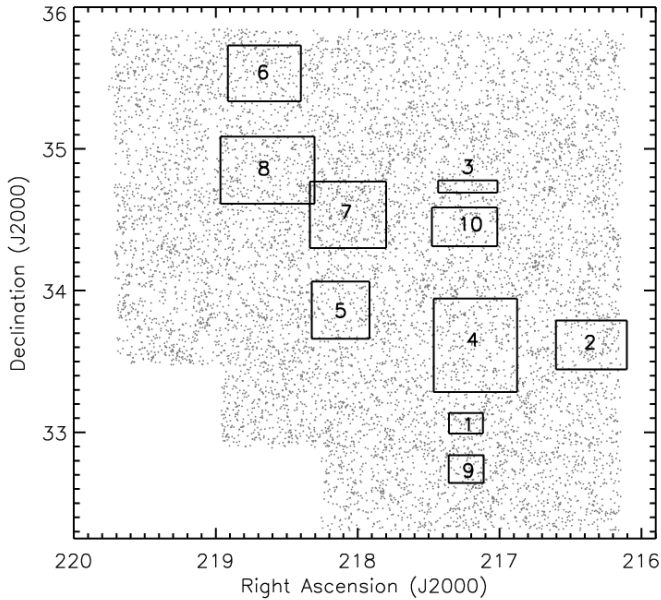}
}
\caption{ \scriptsize Map of the NDWFS survey area with the
coverage of the coadded \zbootes  fields overlayed.  As in Figure
\ref{fig:fullcoverage},
the grey points show the distribution of 25\% of the
 $I<21.5$ extended sources from the NDWFS catalogs.
The number marking
each of the coadded subfields denotes the field number assigned to
that subfield.  The coadded imaging covers 1.24 square degrees of the
7.62 square degree \zbootes region.}
\label{fig:coaddmap}
\end{figure}

The photometric zeropoint for each stacked image was determined by
comparison with photometry publicly available from the SDSS
\citep{am2006}.
Aperture photometry for stars with $18<z<19$ was compared to SDSS
PSF magnitudes.
The mean magnitude offset between the two photometric measurements
was
adopted as the magnitude zeropoint of the field.  In general,
the dispersion
around this median was on the order of $\sigma_z \sim\! 0.03$ mag,
 comparable to the
photometric scatter expected due to errors in the SDSS photometry
\citep{cal}.  As the effective response of the $z$-band filter used
in our
work (including the effects of mirror reflectance and sky absorption)
is
likely different from that of the SDSS system, we examined the
residuals
between the \zbootes photometry and the SDSS photometry as a function
of the
SDSS $i-z$ color of each object.   The photometric residuals showed
no correlation with the object color, and thus no color term has been
applied to the \zbootes photometry.

Several of the \zbootes fields overlapped significantly.
These regions of
significant overlap were coadded, weighted by seeing and
signal-to-noise
in each input frame, to create mosaiced images of the overlap
regions.
Before coadding the individual images in the overlap regions, each of
the input images were background subtracted and the measured counts
per pixel were converted to a true flux density per pixel measurement
using the photometric zeropoint determined from comparison with
public
SDSS photometry discussed above.  Figure \ref{fig:coaddmap}
illustrates
the area of the
\zbootes field included in these coadded observations compared to the
object distribution in the NDWFS optical catalogs.

\section{Source Catalogs}
\subsection{Catalog Generation}
We constructed catalogs for each \zbootes field using Source
Extractor
(SExtractor) version 2.3 \citep{ab1996}.   We detected objects using
a 0.9" FWHM Gaussian convolution kernel and enforced a $3\sigma$
detection threshold.  Pixels were weighted according to the number of
input exposures that contributed to each in order
 to prevent the detections of a large number of spurious
sources around the edges of each field which have fewer average
observations
and thus higher background noise than the centers. 
  For each object detected in
the catalog, we measured the flux in $\sim\!100$ apertures (with
diameter
3", 5", and 7") in a 6 arcminute radius around the object.
We used the
interval containing 68.7\% of the measurements as a measurement
of the
photometric error for each object.  The simulated photometric errors
we calculate from this
method are about a factor of 2 larger than those measured by
SExtractor.

Catalogs were also constructed for each of the stacked images
created for
the overlapping \zbootes fields using the same process as used
for individual
subfields. The final \zbootes catalog was constructed by checking
each
object for duplicate observations.   For any object that was observed
multiply, we define the best observation to be the measurement with
the smallest simulated photometric errors.  The final \zbootes
catalogs
consists of over 200,000 objects.

Objects in the final \zbootes catalogs were matched to detections
in the
NDWFS catalogs (DR3).  For each  NDWFS subfield, we checked for
systematic
offsets in both right ascension and declination between the \zbootes
astrometric system and the NDWFS reference frame.  After removing any
net offset between the NDWFS catalogs and our \zbootes catalogs
(which are
discussed in more detail in \S4.3), the two object lists
were matched with a 1" match tolerance and the name and coordinates
of
the closest match NDWFS detection were recorded in the final
\zbootes catalog.

\begin{figure}[t]
\centering{\includegraphics[angle=0, width=3in]{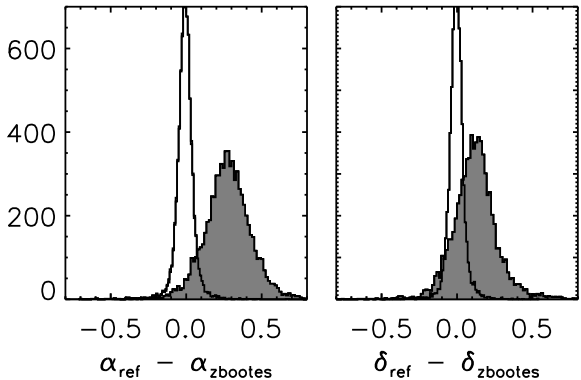}
}
\caption{ \scriptsize Histogram of astrometric offsets (in
arcseconds) between sources reported by \zbootes
and SDSS (unfilled) and \zbootes and NDWFS (filled).  Only
well-detected stars
are used in this plot, so photon noise is not the dominant source of
uncertainty in the astrometry.  In general, SDSS and \zbootes
agree quite
well (50 mas rms).  There are offsets between the \zbootes and
NDWFS for
each of the subfields, resulting in a overall mean offset and
larger dispersion
seen in the comparison between the \zbootes and NDWFS coordinates.  }
\label{fig:astrometry}
\end{figure}
\subsection{Photometric Accuracy}

The final residuals between \zbootes and SDSS photometric
measurements for
well detected stars whose photometry is not affected by non-linearity
in
the 90Prime CCDS are centered around zero with a $1\sigma$
dispersion of
0.035 mag.  The average quoted photometric error for the SDSS
stars is
0.03 mag and thus the majority of the final calibration error in the
\zbootes catalog can be attributed to photometric scatter in the SDSS
photometry.  The remainder of the scatter in the photometric
calibration
is likely due to errors in the large-scale flatfield corrections
to each
frame and imperfect subtraction of the strong fringing the 90Prime
CCDs.

Since 1.25 square degrees were observed more than once in the
$z$Bo\"otes
imaging, we can quantitatively estimate the error in our photometric
measurements, both from calibration errors and reduction
imperfections.
The distribution of fluxes for the $\sim 7200$ stars with $17<z<20$
observed two or more times within \zbootes itself has zero mean
and a $1\sigma$ dispersion of
0.03 mag, in good agreement with the scatter in photometric
calibration
estimated above.

\subsection{Astrometric Precision}

As the \zbootes astrometry was calibrated to the SDSS reference
system,
the agreement between SDSS and \zbootes astrometry is quite good.
Figure \ref{fig:astrometry} shows the differences between SDSS,
\zbootes, and NDWFS astrometry.   The dispersion between the
\zbootes and
SDSS coordinates is $\sim$ 50 milliarcsec (mas) per coordinate while
the agreement with NDWFS is poorer with a dispersion of nearly
130 mas
per coordinate.  SDSS astrometry has a 45 mas dispersion
 per coordinate \citep{pier},
 so
the \zbootes astrometric error is dominated by astrometric errors
in the SDSS
catalogs. Also, notice that the NDWFS and \zbootes astrometric
have systematic offsets in both directions, likely due to the 
different astrometric reference systems used by NDWFS and SDSS imaging.
  Robust comparisons
between
NDWFS and \zbootes (or SDSS) thus require the removal of these
offsets
to properly match objects in each catalog.  Table 1
lists the average shifts between  the astrometry of  each \zbootes
field
and the NDWFS catalog.

\begin{figure}[b]
\centering{\includegraphics[angle=0, width=3.5in]{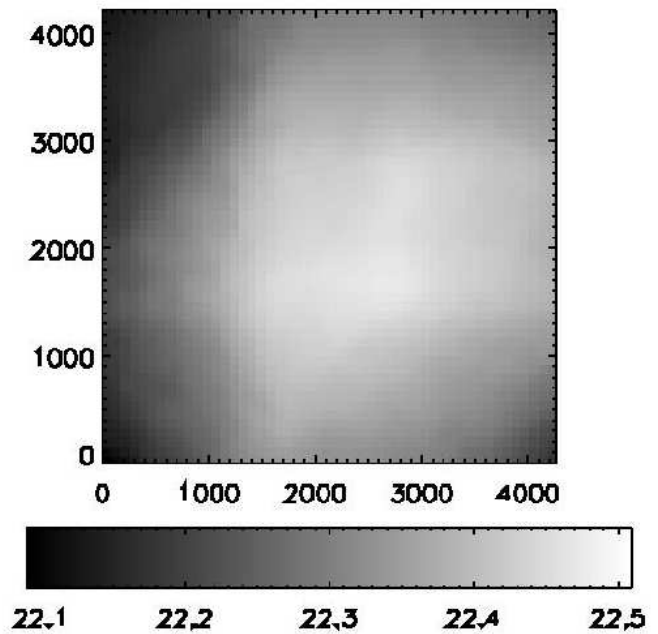}
}
\caption{ \scriptsize Map of the 50\% completeness limit of the b4-1
field in pixel coordinates.  The depth of the catalogs are a strong
function of position in the final mosaiced image.  This is
primarily due
to the decreased number of exposures that were coadded near the
edge of
the mosaiced field.  The completeness can vary as much as 0.5
mag from
the center of the field to the edge.  }
\label{fig:completeness}
\end{figure}

\begin{deluxetable}{ccc|ccc}[b]
\tablecolumns{6}
\label{tab:astrom}
\tablenum{1}
\tablewidth{0pt}
\tabletypesize{\scriptsize}
\tablecaption{Astrometric Offsets Between \zbootes and NDWFS}
\tablehead{
  \colhead{Field Name} &
  \colhead{$\Delta \alpha$ } &
  \colhead{$\Delta \delta$} &
  \colhead{Field Name} &
  \colhead{$\Delta \alpha$} &
  \colhead{$\Delta \delta$} \\
\colhead{} &
\colhead{''} &
\colhead{''} &
\colhead{} &
\colhead{''} &
\colhead{''}}
\startdata
b1-1 & -0.37 & -0.31 & b6-2 & -0.25 & -0.17 \\
b1-2 & -0.30 & -0.37 & b6-3 & -0.31 & -0.12 \\
b1-3 & -0.22 & -0.18 & b6-4 & -0.24 & -0.09 \\
b1-4 & -0.29 & -0.15 & b7-1 & -0.22 & -0.12 \\
b2-1 & -0.30 & -0.05 & b7-2 & -0.26 & -0.10 \\
b2-2 & -0.22 & -0.04 & b7-3 & -0.28 & -0.11 \\
b2-3 & -0.24 & -0.08 & b7-4 & -0.30 & -0.13 \\
b2-4 & -0.21 & -0.09 & b8-1 & -0.35 & -0.10 \\
b3-1 & -0.20 & -0.13 & b8-2 & -0.37 & -0.17 \\
b3-2 & -0.23 & -0.10 & b8-3 & -0.26 & -0.04 \\
b3-3 & -0.25 & -0.15 & b8-4 & -0.33 & -0.12 \\
b3-4 & -0.27 & -0.13 & mos-01 & -0.17 & -0.10 \\
b4-1 & -0.03 & -0.05 & mos-02 & -0.25 & -0.22 \\
b4-2 & -0.01 &  0.01 & mos-03 & -0.21 & -0.10 \\
b4-3 & -0.11 & -0.10 & mos-04 & -0.20 & -0.17 \\
b4-4 & -0.11 & -0.11 & mos-05 & -0.33 & -0.11 \\
b5-1 & -0.17 &  0.02 & mos-06 & -0.21 &  0.07 \\
b5-2 & -0.30 & -0.12 & mos-07 & -0.30 & -0.12 \\
b5-3 & -0.20 &  0.08 & mos-08 & -0.32 & -0.00 \\
b5-4 & -0.28 & -0.18 & mos-09 & -0.13 &  0.04 \\
b6-1 & -0.32 & -0.13 & mos-10 & -0.21 & -0.07 \\
\enddata

\end{deluxetable}

\subsection{Survey Depth}

The depth of the \zbootes images varies between each field and
as a function of position in each field itself due to the variable
number of exposures taken for each field and variable conditions
during the
observations.  In order to
quantify the depth of our catalogs near each detected object,
we added
fake point sources with the same point spread function as measured
from nearby unsaturated stars in the \zbootes images.  For each
field, we perform ten
simulations with each simulation consisting of 3000 fake stars
added to
the coadded frame of each field.  We then record the average 50\%
completeness in a 10
arcminute diameter region around each object detected in our
catalogs.
Figure \ref{fig:completeness} illustrates the variations in the
survey
depth within a single \zbootes field.  As illustrated in the figure,
the variations in survey depth can be as large as 0.5 mags across
the field.

\begin{figure}[b]
\centering{\includegraphics[angle=0, width=3.0in, height=3.0in]{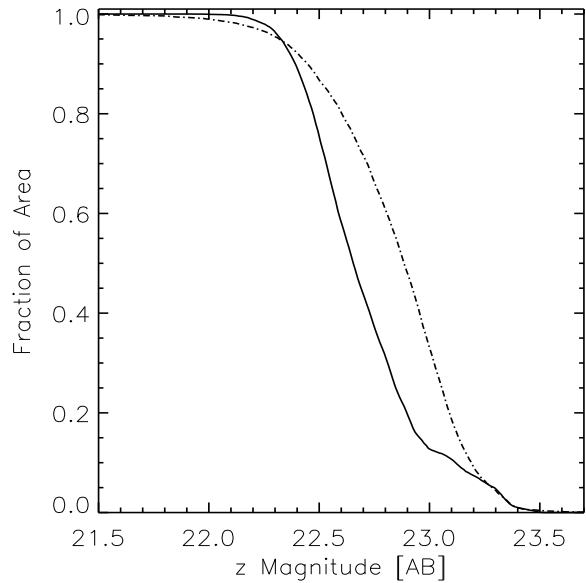}
}
\caption{ \scriptsize Fraction of the \zbootes survey area as
function of the  50\% completeness limit and $3\sigma$ detection
limit of the \zbootes catalogs.  The solid line shows the fraction
of the
\zbootes 7.6 square degree area as a function of the 50\%
completeness
while the dot-dashed line shows the fraction of the survey area
versus the
 $3\sigma$ detection limit in a 3 arcsecond aperture.  The
\zbootes catalog is complete to 22.4 mag over 90\% of the survey
area and 50\%
of the area is complete to 22.7 mag or deeper.  The knee in the
distribution
of 50\% completeness limit near $z=23.0$ is due the deeper limits
present in the
coadded \zbootes fields as discussed in the text.}
\label{fig:compfrac}
\end{figure}

Figure \ref{fig:compfrac} shows the fraction of the \zbootes coverage
area as a function of the 50\% completeness depth and as a function
of the $3\sigma$
detection limits of the catalogs.    The final \zbootes catalog is
50\% complete to
22.4 mag over 90\% of the survey area and 50\% of the survey area is
complete to 22.7 mag.  Thus, the \zbootes catalogs reach more than 2
mag fainter than SDSS over the entire survey region. The mosaiced
images
of the overlapping \zbootes fields reach fainter limits than the
single frames.  Of the 1.24 square degrees covered
in the overlapping fields, 50\%
of the area is complete to 23.3 mag and 90\% is complete to 23.1 mag.
Table 2 lists the average 50\% completeness limit of each of the \zbootes
subfields.   The \zbootes data are thus sensitive to $L^*$ galaxies
to $z\sim1$
and should provide extra constraint on photometric redshifts
measurements
for galaxies at $0<z<1$.
\begin{deluxetable}{cc|cc}
\tablecolumns{6}
\label{tab:depth}
\tablenum{2}
\tablewidth{0pt}
\tabletypesize{\scriptsize}
\tablecaption{Average Completeness Limits (50\%) of \zbootes Fields}
\tablehead{
  \colhead{Field Name} &
  \colhead{Depth} &
  \colhead{Field Name} &
  \colhead{Depth} \\
\colhead{} &
\colhead{mag} &
\colhead{} &
\colhead{mag}}
\startdata
b1-1 & 22.35 & b6-2 & 22.93 \\
b1-2 & 22.50 & b6-3 & 22.95 \\
b1-3 & 22.54 & b6-4 & 22.88 \\
b1-4 & 22.52 & b7-1 & 22.81 \\
b2-1 & 22.59 & b7-2 & 22.91 \\
b2-2 & 22.68 & b7-3 & 22.95 \\
b2-3 & 22.79 & b7-4 & 22.80 \\
b2-4 & 22.72 & b8-1 & 22.63 \\
b3-1 & 22.56 & b8-2 & 22.68 \\
b3-2 & 22.69 & b8-3 & 22.73 \\
b3-3 & 22.67 & b8-4 & 22.69 \\
b3-4 & 22.63 & mos-01 & 22.87 \\
b4-1 & 22.58 & mos-02 & 23.13 \\
b4-2 & 22.50 & mos-03 & 23.05 \\
b4-3 & 22.76 & mos-04 & 23.25 \\
b4-4 & 22.63 & mos-05 & 23.17 \\
b5-1 & 22.39 & mos-06 & 23.35 \\
b5-2 & 22.48 & mos-07 & 23.33 \\
b5-3 & 22.46 & mos-08 & 23.38 \\
b5-4 & 22.41 & mos-09 & 22.84 \\
b6-1 & 22.85 & mos-10 & 23.33 \\
\enddata
\tablecomments{\scriptsize Depth is estimated by the 50\%
completeness
limit of the catalogs.}
\end{deluxetable}
\section{Data Products}

\subsection{Images}
\begin{itemize}

\item Images and weight maps: {\tt [fieldname.fits,
fieldname-weight.fits]} We release the final coadded
images and the associated weight map for each.
The images are in ADU counts per pixel and the magnitude zeropoint
of the
photometry of each image is stored in the {\tt MAGZERO} header
keyword. The weight maps are normalized such that each
pixel reflects the number of exposures that contributed to the image.
All of these
images have world-coordinate system information in the headers.
Note that the orientation of these
images corresponds to the orientation of the 90Prime images
on the sky (north is to the right
and east is upward) and not the standard image orientation.

Regions of the sky that were observed in multiple \zbootes fields
were coadded
as described in \S 3. The resulting images are flux
calibrated and have units of nano-maggies per pixel.  A nanomaggie is
a flux-density unit
equal to $10^{-9}$ of a magnitude zero source.  Since we calibrated
the
\zbootes photometry to SDSS, and SDSS is
nearly an AB system, 1 nanomaggie corresponds to $3.631 \,
\mu \mbox{Jy}$ or $3.631 \times 10^{-29}$ erg s$^{-1}$
cm$^{-2}$ Hz$^{-1}$.  As above, the weight maps associated with each
of the mosaic images reports the total number of 300s exposures that
contributed to each pixel.

\item Photometric Catalogs: [{\tt fieldname-cat.fits}] We also
release binary FITS files
of the
SExtractor catalog for each \zbootes field as well as overlapping
regions.  In
each file, we report
the measured properties of each of the objects detected in the
\zbootes imaging.
A majority of the parameters listed in the catalogs are standard
SExtractor outputs, so we will not repeat the definitions. The
aperture fluxes
and magnitudes reported in the \zbootes catalogs are measured at
12 diameters.  The diameters run from 1" to 10" in steps of 1";
the final two
apertures have diameters of 15" and 20".
The non-standard
parameters included in each of the catalogs are :
\begin{itemize}
\item {\tt NOBS} -- [integer] Mean number of observations (300s
exposures) that contribute to the pixels each object falls on.
Objects
with less than ${\tt NOBS} \lesssim 5$ should be used with caution.
\item {\tt ERR\_[1,3,5,7]} -- [float] Photometric error in a
[1,3,5,7] arcsecond diameter aperture
determined from the dispersion in the local sky background within a 6
arcminute radius around each object.
\item {\tt COMP50} -- [float] The 50\% completeness limit determined
by inserting
$\sim\!30,000$ fake point sources into the images and measuring the
fraction recovered using the same analysis procedure as that
used when
constructing the catalogs.  The local completeness is calculated
within
a 10 arcminute region around each object.
\item {\tt DETECT\_3SIG\_3ARC} -- [float] The local $3\sigma$
detection limit
determined in a 3 arcsecond diameter aperture around each object
based
on measurements of the local variation in the sky background.
\item {\tt PHOTFLAG} -- [integer] For each object, this flag is
set if any of
the pixels contributing to the object detection were in the
non-linear or saturated
regime on the 90Prime CCDs.  Photometry for objects with this
flag set
should be used with caution.
\end{itemize}

\item Final Merged Catalog : [{\tt zbootes-cat.fits}] The final
catalog
represents the merged catalog for the \zbootes imaging.  For objects
included in multiple individual catalogs, the observation with the
smallest photometric error is declared the primary observation and
included in the final catalog. Objects in the final catalog were
cross-matched to the NDWFS optical catalogs using a 1 arcsecond
search
radius.   Before the cross-matching was performed, the locally
determined
astrometric offsets between the \zbootes and NDWFS astrometric
systems
(reported in Table 1) were removed.  The following
parameters are included in the final catalog which are not in
the individual
catalogs:
\begin{itemize}
\item {\tt FIELDNAME} -- [string] Name of the \zbootes field in which
the photometric quantities for each object were measured.
\item {\tt DUPLICATE}  -- [integer] Flag which is set if a given
object
was detected in multiple catalogs.  If an object was detected in
multiple frames,
then the observation with the lowest photometric error was
declared to
be the best and included in the final merged catalog; each object
is listed
in the final catalog only once.
\item {\tt NDWFS\_NAME} -- [string] Name of the nearest NDWFS object
to each
\zbootes detection.  The catalogs were compared with a 1 arcsecond
search
radius; if there were no NDWFS objects within the search radius
of the
\zbootes object, this entry is empty.
\item {\tt NDWFS\_RA} -- [double] Right ascension of the nearest
NDWFS object
to the \zbootes detection in decimal degrees.
\item {\tt NDWFS\_DEC} -- [double] Declination of the nearest
NDWFS object
to the \zbootes detection in decimal degrees.
\end{itemize}
\end{itemize}

\section{Catalog availability}
The \zbootes catalogs described in this paper are
available for download online\footnote[2]{\tt
http://archive.noao.edu/nsa/zbootes.html}.
Any use of these data should include
references to this paper.  If any of the cross-identifications to
the NDWFS catalogs are used,
the appropriate citations should be made to the papers describing
those data.

\section{Acknowledgments}
RJC was funded through a National Science Foundation Graduate
Research
Fellowship.  We are grateful to  Ed Olszewski,  Grant Williams,
and Mike
Lessar for providing the 90Prime instrument and technical information
critical to the reduction and calibration of this data set.  We thank
Buell Jannuzi and Arjun Dey and the NOAO technical staff for hosting
the data provided by this release.  Daniel Eisenstein, Michael Brown,
and Jane Rigby provided many useful comments and suggestions
during the
reduction and calibration of this dataset.

\end{document}